\documentclass[a4paper,twoside]{article}
\usepackage[utf8]{inputenc}
\usepackage[T1]{fontenc}
\usepackage{amsmath}
\usepackage{amssymb}
\usepackage{amsthm}
\usepackage{tsc}
\usetikzlibrary{fit, angles, arrows.meta}
\usepackage{subcaption}
\usepackage{hyperref}
\usepackage[linesnumbered]{algorithm2e}
\usepackage{mathtools}
\usepackage[squaren,binary]{SIunits}

\usepackage{epsfig}
\usepackage{calc}
\usepackage{multicol}
\usepackage{pslatex}
\usepackage{apalike}
\usepackage[bottom]{footmisc}

\usepackage{SCITEPRESS}     

\newcommand\BC{{\mathsf{BC}}}
\newcommand\WM{{\mathsf{WM}}}
\newcommand\TSC{{\mathsf{TSC}}}
\newcommand\Traj{{\mathbf{Traj}}}

\newcommand\HFC{{\mathsf{HFC}}}

\newcommand\TSCs{{\mathbb{TSC}}}

\newcommand\AandB{\text{\inlinetsc{\tscAnd{\item{\subchart{A}}\item{\subchart{B}}}}}}
\newcommand\AorB{\text{\inlinetsc{\tscOr{\item{\subchart{A}}\item{\subchart{B}}}}}}
\newcommand\AseqB{\text{\inlinetsc{\subchart{A}\tscSeq\subchart{B}}}}

\newcommand\started{{\mathsf{started}}}
\newcommand\complete{{\mathsf{complete}}}
\newcommand\ok{{\mathsf{ok}}}

\def\ltlpexp#1#2\endp{\def\ltlsecond{#2}\ifx\ltlsecond\empty#1\else#1\rightarrow#2\fi}

\newcommand\ctrue{\textsf{true}}

\newcommand\tuple[1]{\langle#1\rangle}
\newcommand\CheckSatN{\textsc{CheckSatN}}
\newcommand\CheckSatS{\textsc{CheckSatS}}
\newcommand\Sat{\textsc{Sat}}

\newcommand\xmin{{\ensuremath{\underline{\mathtt{x}}}}}
\newcommand\xmax{{\ensuremath{\bar{\mathtt{x}}}}}
\newcommand\ymin{{\ensuremath{\underline{\mathtt{y}}}}}
\newcommand\ymax{{\ensuremath{\bar{\mathtt{y}}}}}
\newcommand\carI{{\ensuremath{\mathtt{carI}}}}
\newcommand\carJ{{\ensuremath{\mathtt{carJ}}}}
\newcommand\lLane{{\ensuremath{\mathtt{lLane}}}}
\newcommand\rLane{{\ensuremath{\mathtt{rLane}}}}

\SetKw{Not}{not}
\SetKw{Or}{or}
\SetKw{Continue}{continue}
\SetKw{With}{with}

\newtheorem{example}{Example}
\newtheorem{definition}{Definition}

\begin{document}
	\title{A Consistency Analysis Method for Traffic Sequence Charts}
	\author{\authorname{Jan Steffen Becker}
		\affiliation{German Aerospace Center (DLR) e.V. \\ Institute of Systems Engineering for Future
		Mobility \\ Oldenburg, Germany}
		\email{jan.becker@dlr.de}}
	
	\onecolumn \maketitle \normalsize \setcounter{footnote}{0} \vfill
	
\begin{abstract}
	The trend in the development of highly automated vehicles goes towards scenario-based methods. Traffic Sequence Charts are a visual but yet formal language for describing scenario-based requirements on highly automated vehicles. This work presents an approach for finding inconsistencies (conflicts) in a set of scenario-based requirements formalized with Traffic Sequence Charts. The proposed method utilizes satisfiability modulo theories solving on two-sided approximations of possible vehicle behavior. This ensures that found inconsistencies are not caused by approximations, but also occur when applying exact methods. Applicability and scalability of the analysis technique is evaluated in a case study. 
	\paragraph*{keywords} Scenario-based Development, Traffic Sequence Charts, Consistency, ISO~26262, Satisfiability Modulo Theories, Bounded Model Checking 
\end{abstract}

	\section{Introduction}

The development of highly automated vehicles (HAVs) has to deal with highly complex environments. In SAE Level 3 \cite{SAE3016} and above, the automated vehicle is expected to drive safely within its design domain. The challenge of ensuring provably safe behavior starts already in the early phases of the development process. Potentially critical driving situations have to be identified, and strategies need to be defined that lower risks to an acceptable minimum \cite{Kramer2020,Neurohr2021}. The number and complexity of different situations is enormous as it includes both static and dynamic aspects, such as road and weather conditions, different traffic participants, and driving maneuvers \cite{koopman2019many}. Therefore, the trend goes towards scenario-based approaches that aim at clustering this unmanageable set of different situations and scenarios into a manageable set of scenario classes \cite{Kalisvaart2020,Menzel2018,Riedmaier2020}. During the whole development process, scenario descriptions of different granularity are used. For example, identified critical scenario classes are characterized by textual scenario descriptions, so-called \emph{functional scenarios} during the concept phase, while testing requires \emph{concrete scenarios} that allow an exact reproduction in a (virtual or physical) test bed \cite{Menzel2018}. 

This rigorous scenario-driven approach calls for specifying also the behavioral requirements on the system, i.e., how the HAV shall behave in certain situations, in a scenario-based manner. Traffic Sequence Charts (TSCs) \cite{Damm2018} are a graphical formalism that enables exactly this. In a TSC, traffic situations are graphically depicted and assembled to formalized versions of functional scenarios, called \emph{abstract scenarios} \cite{Neurohr2021}. With the same technique, TSCs also allow to formalize requirements. 

As a research contribution to the field of scenario-based development of HAV, this paper reports on the author's doctoral thesis entitled ``A Consistency Analysis Method for Traffic Sequence Charts'' that is currently under work. In short, the thesis aims on the development and experimental evaluation of an automated consistency analysis approach for TSCs. 

The remainder of the paper is structured as follows. Section~\ref{sec:problem} presents the research problem and objectives of the thesis. Section~\ref{sec:related-work} presents background work, followed by an introduction to TSCs in Section~\ref{sec:tscs}. Section~\ref{sec:tscs} also introduces the running example. In Section~\ref{sec:methods}, key ideas of the methodology and expected results are named. The proposed consistency analysis method is then described in more detail in Section~\ref{sec:consistency}. Some experimental evaluation results are presented in Section~\ref{sec:evaluation}. Finally, Section~\ref{sec:conclusion} concludes the paper with a short discussion and outlook to future work.

\section{Research Problem and Objectives}\label{sec:problem}

Good systems engineering practice mandates that requirements shall comply to the three big Cs \emph{correctness}, \emph{completeness}, and \emph{consistency} \cite{ZowghiGervasi03,kamalrudin2015review}. Correctness is usually defined as the combination of completeness and consistency \cite{ZowghiGervasi03}. Normative standards -- such as ISO~26262 \cite{ISO26262} in the automotive domain -- state completeness and consistency as mandatory properties of any requirements specification. Completeness is quite hard to grasp, and only be defined relatively, for example with respect to the outcome of a preceding hazard analysis \cite{leveson2000completeness}. Consistency means that a set of requirements is free of contradictions. This includes both contradictions within itself and with respect to other requirements \cite{ISO26262}. Detecting inconsistencies early in the development process avoids implementation faults and saves time and money \cite{Feiler2010}. 

A lot of different approaches (e.g. \cite{jaffe1991,EllenSieverdingHungar14,Becker2018,FilipovikjRNS17,AichernigHLNT15,PostHP11}) exist that formally define consistency, but none of it is directly applicable to TSCs. Therefore, the presented PhD thesis addresses two major research questions
\begin{enumerate}
	\item How can consistency for traffic sequence charts be defined formally?
	\item How can a consistency check that uses these definitions of consistency be automated?
\end{enumerate}

The research objectives of the thesis can be summarized as follows. 
\begin{enumerate}
	\item Develop adequate consistency notions for TSCs
	\item Define a decision procedure for the consistency notions
	\item Formally prove correctness of the encoding
	\item Prototypically implement the approach
	\item Evaluate effectiveness and scalability of the approach
\end{enumerate}
The experiments for evaluation are further described in Section~\ref{sec:evaluation}.

	\section{State of Practice}\label{sec:related-work}
Consistency analysis for TSCs bridges research from different fields. 

In the automotive industry, scenario-based development \cite{Kalisvaart2020,Menzel2018,Riedmaier2020} is the answer to the growing complexity of HAV. Menzel et. al. \cite{Menzel2018} identified a wide range of applications of scenario descriptions in a traditional development process. As already mentioned in the introduction, scenarios are used in different abstraction levels. Traffic sequence charts \cite{DammKMPR17,Damm2018} have been developed to formally describe scenarios with a high abstraction level. TSCs have been used in a wide range of applications, e.g. scenario mining \cite{DammMR19a}, test design \cite{DammMR19a} and runtime monitoring \cite{Grundt2022}. In contrast, OpenSCENARIO\footnote{\url{https://www.asam.net/standards/detail/openscenario-xml/}, accessed on 2024-02-28. OpenSCENARIO XML was formerly known as OpenSCENARIO 1.x} XML has been developed as an industrial standard for describing automotive scenarios on a lower abstraction level. OpenSCENARIO mainly targets simulation of scenarios and is supported by industrial-grade driving simulators. There is also an approach to generate OpenSCENARIO XML specifications from TSCs \cite{Becker2022}. Another ASAM standard, OpenSCENARIO DSL\footnote{\url{https://www.asam.net/standards/detail/openscenario-dsl/}, accessed on 2024-02-28. OpenSCENARIO DSL was formerly known as OpenSCENARIO 2.0}, allows to specify abstract scenarios, but there is no complete tool support for these new features, yet.

In the ENABLE-S3 research project \cite{ENABLES3SummaryResults2019}, a reference process for scenario-based development has been defined. Here, consistency analysis is one step in the \emph{requirement and scenario elicitation} activity, but is not  discussed further. However, automated consistency analysis has already been developed for other types of formal requirement specifications. For example, the software cost reduction (SCR) toolset \cite{jaffe1991,heitmeyer1996automated} provides an automated consistency analysis for automata-like specifications. More recent work \cite{EllenSieverdingHungar14,Becker2018,FilipovikjRNS17,AichernigHLNT15,PostHP11} considers pattern-based requirements. 

In most of the aforementioned work, satisfiability modulo theories (SMT) solving in connection with bounded model checking (BMC) is used to find inconsistencies. All these works consider discrete transition systems. Consistency is then reduced to existence of system runs that satisfy the requirements. SMT is well suited for consistency analysis because it is capable of both synthetizing system runs (which are a witness for consistency) and proving non-existence of satisfying runs (which proves inconsistency). 
In the authors earlier work \cite{Becke20}, this idea is already applied to TSCs. Instead of discrete transition systems, the translation process to SMT is based on earlier work about duration calculus, which has some similarities to TSCs. However, it over-approximates possible vehicle behavior which means that correctness of witness trajectories cannot be guaranteed. The aforementioned generation of OpenSCENARIO from TSCs \cite{Becker2022} is based on the same method, but extends it with a simple vehicle dynamics model. As a consequence, generated concrete scenarios are correct, but not all possible scenarios can be found. 
A similar technique is also applied by Eggers et al. \cite{EggerSTBB18} to scenario specifications that are similar to existential TSCs, but more restrictive. To the author's knowledge, these are the only works that tackle consistency of traffic scenario specifications. 

The present work builds upon the existing work on TSC consistency analysis \cite{Becke20} and extends it with the TSC instantiation technique \cite{Becker2022} that has already been used for OpenSCENARIO generation. The latter uses a conservative linear approximation of vehicle dynamics in conjunction with Bézier spline trajectory planning. Alternatives would be statespace exploration techniques for hybrid systems \cite{henzinger1997hytech,frehse2005phaver,frehse2011spaceex,chen2016,eggers2011}. However, these numerical approaches focus on ordinary differential equations without a known closed form solution and unfortunately scale badly with high dimensional state spaces. Plaku et al. \cite{plaku2007hybrid,plaku2013falsification} overcome these limitations by combining searches on a discrete state spaces to guide exact simulations. The approach depends on a discrete approximation of the reachable state space which needs to be provided manually. Therefore, it is not suited for an automated consistency analysis.

	\section{Traffic Sequence Charts}\label{sec:tscs}
\begin{figure*}[t]
	\centering
	\subcaptionbox{Sequence\label{fig:seqchart}}[.12\textwidth]{$\AseqB$}%
	\subcaptionbox{Choice\label{fig:choice}}[.12\textwidth]{$\AorB$}%
	\subcaptionbox{Concurrency\label{fig:concurrency}}[.12\textwidth]{$\AandB$}%
	\subcaptionbox{Timing annotations\label{fig:timepins}}[.24\textwidth]{\inlinetsc{\tscAnd{\item{\subchart{A}\tscSeq\subchart{B}\tscTimeAnnot{-}{\tscPin{p}\hfill\tscPin{q}}\tscSeq\subchart{C}}\item{\subchart{D}\tscXSeq{\tscPin{p}}\subchart{E}\tscXSeq{\tscPin{q}}\subchart{F}}}\tscTimeAnnot{*-*}{\tscHGS{d}\hfill\tscHGE{d<10\mathrm{s}}}}}
	\subcaptionbox{Empty invariant\label{fig:emptyinv}}[.12\textwidth]{\inlinetsc{\tscTrue}}
	\subcaptionbox{Invariant node\label{fig:invariant}}[.24\textwidth]{\inlinetsc{\snapshot{\includegraphics[scale=.5]{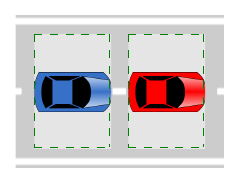}}}}
	\caption{Syntax of basic charts}\label{fig:basic-charts}
\end{figure*}

TSCs have been developed with the goal of creating a description language that connects the intuitiveness of depicting traffic situations graphically with well-defined semantics. The core concept of a TSC is a so-called \emph{invariant node} that graphically depicts a traffic situation (or, as we will see later, a combination of situations). Here, symbols stand for objects, and their placement indicates spatial relations. Invariant nodes are assembled to \emph{basic charts} by combining them using the operators \emph{sequence}, \emph{choice}, and \emph{concurrency} depicted in Figure~\ref{fig:basic-charts}. The operators can be arbitrarily nested. Furthermore, it is possible to add timing annotations (Figure~\ref{fig:timepins}) to basic charts---time pins (\tscPin{}) with the same label synchronize time points, and hour glasses (\tscHGS{}, \tscHGE{}) express duration constraints. The full TSC formalism \cite{DammMPR18} also allows negation of basic charts, which has been omitted for this paper. Experience shows that negation is seldom needed in practice because it allows almost any (also unintended) behavior. So, this is only a minor limitation.    
 
\begin{figure*}[t]
	\centering
	\includegraphics{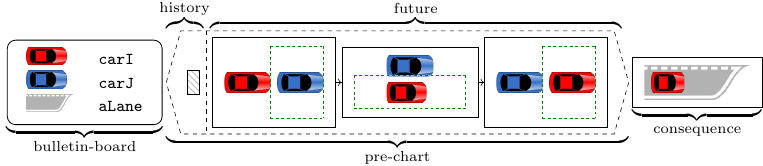}
	\caption{A TSC and its parts. The \emph{bulletin-board} declares object symbols with global scope in the TSC; the \emph{pre-chart} is a triggering condition for the TSC, where \emph{history} describes past behavior; the \emph{consequence} is the requirement obligation that shall be maintained during the \emph{future} behavior.}\label{fig:tsc-parts}
\end{figure*}

A \emph{requirement TSC} resembles a typical specification pattern, sometimes called \emph{response property}\footnote{With the full TSC language specified in \cite{DammMPR18} also a wide range of other patterns can be realized.} \cite{dwyer1999patterns,konrad2005real}. It consists of three parts depicted in Figure~\ref{fig:tsc-parts}: a \emph{bulletin board} declaring symbols referring to global object variables, a \emph{pre-chart} describing a triggering condition split into history (left part) and future (right part), and the \emph{consequence} defining a reaction to the trigger that is synced with the future. It expresses that whenever the pre-chart is observed, then also the consequence shall be observed; thereby, the consequence has to happen in parallel to the future. For simplicity, we denote a requirement TSC by a triple $\tuple{H, F, C}$ made of history $H$, future $F$, and consequence $C$.  

\subsection{Interpretation of Spatial Views}

TSCs are always interpreted with respect to a \emph{world model} and a \emph{symbol dictionary}. The world model defines the domain ontology for the specification. At least, it defines the \emph{object types} that a TSC may speak about together with the attributes. There is no unique way of defining a world model. For example, one could define the world model in terms of a UML class diagram \cite{Booch2005} or description logic \cite{BaaderCMNP07}. Earlier work on TSCs \cite{DammMPR18} sees the world model as a network of communicating hybrid automata.  

\begin{figure}[bt]
	\centering
	\includegraphics[width=.8\columnwidth]{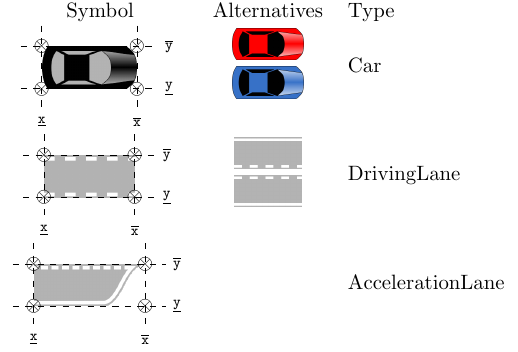}
	\caption{Symbol dictionary for the running examples\label{fig:sdict}}
\end{figure}

The \emph{symbol dictionary} (Figure~\ref{fig:sdict}) defines the symbols that are used to represent objects from the world model within spatial views. This way, it provides the link between the world model and the TSC. Each object symbol has a type and a set of \emph{anchors}. Anchors bind selected points of a symbol to positions in a 2D space. In this paper, the anchor points are always placed in the four corners of a symbol and describe the object bounding boxes in our global coordinate system. So, the bottom-left anchor binds to $(\xmin, \ymin)$ and the top-right anchor to $(\xmax, \ymax)$ in object attributes. The alternative symbol variants (second column of Figure~\ref{fig:sdict}) are used to make bulletin board symbols visually distinguishable (e.g., $\carI$ from $\carJ$ in Figure~\ref{fig:tsc-parts}).

Now, we come to spatial views. As said above, symbols declared in the bulletin board stand for objects of the corresponding type. \emph{Somewhere boxes} (green dashed rectangles) and \emph{nowhere boxes} (red crossed rectangles) are special symbols that structure a spatial view into a hierarchy of frames---the spatial view spans the top-level frame and each somewhere or nowhere box spans an inner frame. The anchors of symbols and boxes are used to define spatial relations between the objects. 
In each frame, the left-to-right and bottom-to-top orderings of anchors induce an ordering of corresponding positions along $x$ and $y$ axes of the global coordinate system. This creates implicit spatial relations between objects directly contained in the same frame. Hence, the traffic situation depicted in a somewhere box takes place \emph{somewhere} within the region spanned by the frame. Explicit spatial relations are defined by so-called \emph{distance arrows} that constrain distances between anchors in $x$ or $y$ direction. Additional constraints over the object attributes are displayed as a textual label connected to the object symbols.  

The following examples show how spatial views can be translated to mathematical formulae. The algorithm to construct the formulae can be found in \cite{DammMPR18}. 

\begin{figure}[t]
	\subcaptionbox{bulletin-board\label{fig:ex-bb}}{\tscHeader{
	\includegraphics[height=1em]{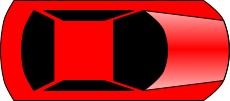} & carI \\
	\includegraphics[height=1em]{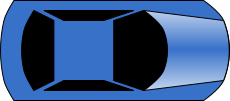} & carJ \\
	\includegraphics[height=1em]{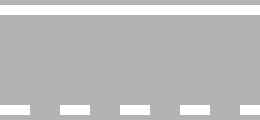} & lLane \\
	\includegraphics[height=1em]{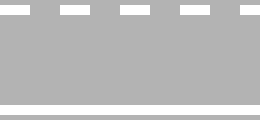} & rLane 
}}\hfill
	\subcaptionbox{SV~1\label{fig:sv-between-lanes}}{\includegraphics[scale=.5]{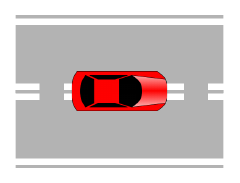}}\hfill
	\subcaptionbox{SV~2\label{fig:sv-distance}}{\includegraphics[scale=.5]{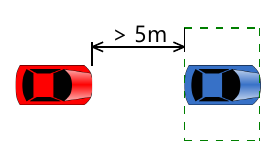}}\hfill
	\subcaptionbox{SV~3\label{fig:sv-nowhere}}{\includegraphics[scale=.5]{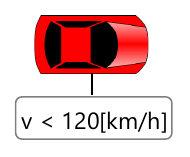}}\hfill
	\subcaptionbox{SV~4\label{fig:sv-predicate}}{\includegraphics[scale=.5]{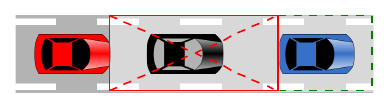}}\hfill
	\caption{Spatial views from Example~\ref{ex:sv-semantics}}\label{fig:example-sspatial-views}
\end{figure}
\begin{example}\label{ex:sv-semantics}
	The following is the semantics of the spatial views in Figure~\ref{fig:example-sspatial-views} when using the bulletin-board in Figure~\ref{fig:ex-bb}. 
	\begin{description}
		\item[SV~1] expresses that \carI{} crosses the border between the left and the right lanes. 
		\[\begin{array}{ccl}
			&&\rLane.\xmin=\lLane.\xmin \\
			&<&\carI.\xmin < \carI.\xmax \\
			&<&\rLane.\xmax=\lLane.\xmax \\
			\wedge && \rLane.\ymin<\carI.\ymin \\
			&<& \rLane.\ymax = \lLane.\ymin \\
			&<& \carI.\ymax < \lLane.\ymax
		\end{array}\]  
	\item[SV~2] expresses that \carI{} is more than 5m behind \carJ{}. Note that because of the somewhere box around \carJ{} only spatial relations in $x$-directions are evaluated\footnote{Because the left and right borders of \carJ{} and the somewhere box are aligned, \carJ{} can be located on the $y$-axis arbitrarily within the box, but not on the $x$-axis.}.  
	\[\carJ.\xmin - \carI.\xmax > \unit{5}{\meter}\]
	\item[SV~3] shows a textual annotation that constraints the speed of \carI{}:  
	\[\carI.\mathtt{v} < \unit{120}{\kilo\meter\per\hour}\]\break
	\item[SV~4] requires that there is no other car $\mathtt{c}$ between \carI{} and \carJ{} on some lane $\mathtt{l}$:  
	\begin{align*}
		\MoveEqLeft \exists \mathtt{l} \in \mathtt{Lane}: \\
		&\quad\mathtt{l}.\xmin < \carI.\xmin < \carI.\xmax \\
		&\qquad< \carJ.\xmin < \carJ.\xmax < \mathtt{l}.\xmax \\
		&\wedge \mathtt{l}.\ymin < \carI.\ymin < \carI.\ymax < \mathtt{l}.\ymax \\
		&\wedge \mathtt{l}.\ymin < \carJ.\ymin < \carJ.\ymax < \mathtt{l}.\ymax \\
		&\wedge \nexists \mathtt{c}\in \mathtt{Car}: \carI.\xmax < \mathtt{c}.\xmin < \mathtt{c}.\xmax < \carJ.\xmin \\
		&\qquad \wedge \mathtt{l}.\ymin < \mathtt{c}.\ymin < \mathtt{c}.\ymax < \mathtt{l}.\ymax
	\end{align*}
	The object variables $\mathtt{l}$ and $\mathtt{c}$ are existentially quantified because the corresponding symbols are not contained in the bulletin board. 
	\end{description}
\end{example}

\subsection{Chart Semantics}\label{sec:chart-semantics}
On top of the semantics of spatial views we define the semantics of charts. \emph{Satisfaction} of a chart is defined with respect to a concrete \emph{trajectory}. 
\begin{definition}
	A \emph{trajectory} over a set $O$ of objects (each having some type from the world model) is a function 
	\[\pi: \mathbb{R}_{\geq 0} \to \mathcal{U}^{\mathcal{A}(O)}\]
	that assigns, for any point $t \in \mathbb{R}_{\geq 0}$ in time, a value to any attribute $o.a \in \mathcal{A}(O)$ of any object $o \in O$. Here, $\mathcal{U}$ is the universe of values used by the world model (e.g., reals and Booleans).
\end{definition}
Given some trajectory, we can evaluate a spatial view at any point in time using the derived formula (given correct typing and that the global objects from the bulletin board are present in the trajectory). A spatial view is \emph{satisfied} at time $t$ if the formula evaluates to $\ctrue$ under the evaluation of all object attributes (including derived ones) given by $\pi(t)$.  

Satisfaction of a basic chart is always defined with respect to an interval $[b, e] \subseteq \mathbb{R}_{\geq 0}$. Furthermore, a time value $t_p$ is selected for every time pin $p$. 
\begin{definition}\label{def:chart-semantics}
	Given some trajectory and time values $t_p$ for time pins $p$, a basic chart is satisfied on some interval $[b, e]$ if the following holds.
	\begin{itemize}
		\item Invariant nodes: $b < e$ and the spatial view holds for all $t \in [b, e)$. 
		\item Empty invariant node: $b < e$ 
		\item Sequences (Figure~\ref{fig:seqchart}): there exists some $m \in [b, e]$ such that $A$ is satisfied on $[b,m]$ and $B$ on $[m,e]$.
		\item Choices (Figure~\ref{fig:choice}): $A$ or $B$ is satisfied on $[b, e]$.
		\item Concurrency (Figure~\ref{fig:concurrency}): both $A$ and $B$ are satisfied on $[b, e]$.
	\end{itemize}
	For charts with timing annotations, the following has to hold additionally.
	\begin{itemize}
		\item Sequences with a time pin $p$ require $m=t_p$.
		\item Charts with a sequence $p_1, \dots, p_n$ of time pins require $b \leq t_{p_1} \leq \dots \leq t_{p_n} \leq e$.
		\item Charts with an hour glass labeled with a free variable $d$ and a constraint $\psi(d)$ over $d$ require that $\psi(e-b)$  evaluates to $\ctrue$ (i.e., when replacing $d$ by $e-b$). 
	\end{itemize}
A requirement TSC is \emph{satisfied} on a trajectory, if for all $b \leq m \leq e \in \mathbb{R}_{\geq 0}$ holds: whenever there are time pin values such that the history is satisfied on $[b,m]$ and the future is satisfied on $[m,e]$, then there are time pin values such that the consequence is satisfied on $[m,e]$. 
\end{definition} 
Note that for invariant nodes, the end point $e$ is explicitly excluded from the interval. This allows non-overlapping sequences of invariants. 

\begin{example}
	The chart in Figure~\ref{fig:timepins} is satisfied on an interval $[b,e]$ if $e - b < 10\mathrm{s}$ and there are time points $m_1, m_2, m_3, m_4$ such that 
	subcharts $A$, $B$, $C$, $D$, $E$, and $F$ are satisfied on $[b, m_1]$, $[m_1, m_2]$, $[m_2, e]$, $[b, m_3]$, $[m_3, m_4]$, and $[m_4, e]$ respectively, and
	\begin{gather*}
		b \leq m_1 \leq t_p < t_q \leq m_2 \leq e \\
		b \leq m_3 = t_p \leq m_4 = t_q \leq e \enspace.
	\end{gather*}
Because time pins are existentially quantified, this is equivalent to 
\[b \leq m_1 \leq m_3 \leq m_4 \leq m_2 \leq e \enspace.\]
\end{example}

In Section~\ref{sec:defining-consistency}, consistency is reduced to satisfiablility of TSCs. Satisfiability asks whether there exists at least one trajectory that satisfies the TSC. Therefore, we assume that a world model $\WM$, beneath a type hierarchy, defines the universe $\Traj(\WM)$ of all possible trajectories. In other words, it describes all possible behavior. The concrete world model used throughout this paper is introduced later on in Section~\ref{sec:worldmodel}

\begin{definition}
	A basic chart $\BC$ is \emph{satisfiable} in a world model $\WM$, written $\Sat_{\WM}(\BC)$, if there exists a a trajectory $\pi \in \Traj(\WM)$, a time point $e > 0$ and time pin values such that $\BC$ is satisfied on $\pi$ on $[0, e]$. 	  
	
	A TSC $\TSC$ is \emph{satisfiable} in a world model $\WM$, written $\Sat_{\WM}(\TSC)$ if there exists a trajectory $\pi \in \Traj(\WM)$ such that $\TSC$ is satisfied on $\pi$. 
\end{definition}

\subsection{Use Case World Model}\label{sec:worldmodel}

\begin{figure}[bt]
	\centering
	\includegraphics[scale=0.8]{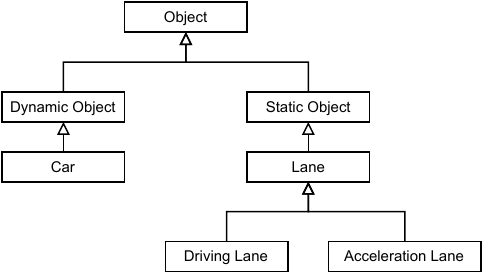}
	\caption{The world model that is used in the use case. Each box represents an object type, arrows denote inheritance\label{fig:worldmodel}}
\end{figure}

Figure~\ref{fig:worldmodel} shows an excerpt of the world model used during this paper\footnote{It is used by both the examples and the evaulation case.}. Every object type defines a set of attributes and invariants. The attributes include positions in form of the minimum $(\xmin, \ymin)$ and maximum $(\xmax, \ymax)$ coordinates of the object's bounding box. We use a road coordinate system where the reference line of the rightmost lane is chosen as the $x$-axis. Lane boundaries are then expressed in terms of $\mathtt{start}$, $\mathtt{length}$, and $\mathtt{width}$ of the lane. This does not mean that all roads symbolized in a TSC are straight roads in reality. Coordinate transformations such as Lanelet transformation \cite{bender2014lanelets} allow to interpret spatial views on curved roads, too. 

\begin{figure}
	\scalebox{.75}{\input{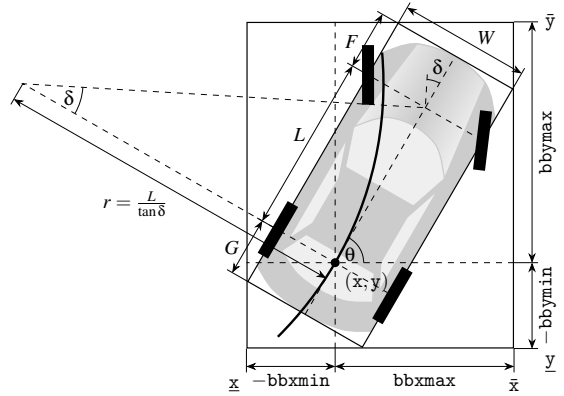}}
	\caption{Bounding box for a car and illustration of the single track model (figure taken from \cite{Becker2024}). Constants $F$, $G$, $L$ and $W$ together describe the vehicle dimensions. The turning radius $r$ depends on wheel base $L$ and Ackermann steering angle $\delta$.\label{fig:bounding-box}}
\end{figure}

The bounding box of a car is expressed by offsets $\mathtt{bb(x|y)(min|max)}$ to the reference point as denoted in Figure~\ref{fig:bounding-box}. 

Cars move according to a single track model given by
\begin{align*}
	\dot{\mathtt{x}}&= \mathtt{v}\cos\theta &
	\dot{\mathtt{y}} &= \mathtt{v}\sin\theta &
	\dot\theta &= \frac{\mathtt{v}}{r}=\frac{\tan \delta}{L}\mathtt{v} &
	\dot{\mathtt{v}} &= \mathtt{a} 
\end{align*}
subject to the constraints $|\delta| \leq \delta_{\max}$ and lateral acceleration $|a_{lat}| = |\mathtt{v}\dot\theta| \leq 0.4g$. 
The lateral acceleration bound of $0.4g\approx\unit{3.92}{\meter\per\squaren\second}$ is stated in the literature \cite{Schramm2014} as a validity constraint for the single track model.

\subsection{Satisfiability Modulo Theories and Bounded Model Checking}

Satisfiability modulo theories (SMT) solving is an extension of classical satisfiability (SAT) solving that allows to combine different theories, e.g., linear arithmetics and bit vectors. An overview on SMT solving can be found in \cite{BarrettSST09}. An SMT solver (supporting a set of theories) takes as an input a problem description in form of a set $\Phi$ of constraints. The solver then tries to either find a satisfying assignment (returning \texttt{sat}) or prove unsatisfiability (returning \texttt{unsat}) of $\Phi$ within the used theories. If the solver cannot do either (e.g., because of a timeout or incompleteness of the implemented decision procedure), it returns \texttt{unknown}. 

With bounded model checking (BMC) \cite{Armando2009}, SMT solving is utilized to check bounded reachability in symbolic transition systems. In BMC, the state vector is encoded in a set $\mathbf{X}$ of variables. A BMC problem is a triple $(I, T, F)$ of constraints over $\mathbf{X}$. The constraints $I(\mathbf{X})$ and $F(\mathbf{X})$ characterize the initial and final states of the system. The possible transitions are encoded in the constraint $T(\mathbf{X}, \mathbf{X}')$ that evaluates to $\ctrue$ whenever there is a transition between the current state $\mathbf{X}$ and the next state $\mathbf{X}'$. Introducing an instance $\mathbf{X}_i$ of $\mathbf{X}$ for every step $i=0, \dots, n$, the constraint
\[ I(\mathbf{X}_0) \wedge \bigwedge_{i=1}^n T(\mathbf{X}_{i-1}, \mathbf{X}_i) \wedge F(\mathbf{X}_n)\]
characterizes all accepting runs with $n$ steps. 

	\section{Methodology and Expected Results}\label{sec:methods}
	
The consistency analysis shall be based on a rigorous formal method. Due to the incompleteness of scenario descriptions and domain models in early phases of the development process, validity of the found inconsistencies is mote important than completeness of the findings. Following some of the existing approaches \cite{EllenSieverdingHungar14,Becker2018,FilipovikjRNS17}, it seems a good idea to base a consistency notion for TSCs on the existence of satisfying trajectories. Because TSCs describe requirements as sequences of invariants, SMT solving seems a promising approach to tackle the problem. Some of the related work \cite{EllenSieverdingHungar14,Becker2018,FilipovikjRNS17,EggerSTBB18} also uses SMT solving for the generation of trajectories. 

The consistency analysis shall provide feedback about a scenario specification already in early phases of the development process. Here, scenario specifications may be still incomplete. Furthermore, the operation environment and physical constraints of the HAV may be under-specified. Because the consistency analysis has to deal with large sets of requirements in a potentially underspecified context, performance shall be favored over completeness of the results. 
TSCs describe scenarios in a dense time domain, and require to consider non-linear movement given usually as differential equations or hybrid automata \cite{Damm2018}. Numerical approaches to explore this kind of continuous-time hybrid systems exist \cite{henzinger1997hytech,frehse2005phaver,frehse2011spaceex,chen2016,eggers2011}, but scalability of these methods is an issue. The TSC semantics adds another layer of complexity. Results on Duration Calculus (that has similar operators to TSCs) \cite{ChaochenHS93,Bouajjani1995} show that this might easily lead to state explosion. Therefore, the consistency analysis uses two-sided approximations. As explained in detail in Section~\ref{sec:decision-procedure}, both a necessary and a sufficient condition for the existence of trajectories is developed. The former produces discrete approximations that don't ensure validity (except from continuity). The latter is incomplete, but produces trajectories that are valid with respect to the TSCs and a realistic vehicle dynamics model. Because the considered requirement specification mainly address maneuver and trajectory planning, a single-track model (explained in Section~\ref{sec:worldmodel}) is sufficient. For reasons of efficiency, the methods are designed to be usable with solvers for mixed Boolean and linear real arithmetic. 

The developed prototype shall be applied to medium sized (sets of) TSCs, thereby evaluating scalability (i.e., in terms of runtime) and completeness of the approach. It is expected that the experimental evaluation shows that an SMT-based automated consistency analysis for TSCs is practicable. By the nature of SMT solving, it must be expected that the execution time for solving the generated SMT problems is expnential to the size of the TSCs. However, it shall be possible to design the analysis method in a way the overall number of SMT problems to be solved in practice grows only polynomial to the size of the specification, and that each SMT problem is small enough to produce a result in acceptable time. Furthermore, it is expected that the experiments show some limitations of the approach wrt. completeness of found inconsistencies, but that the chosen approximations are sufficient to find common specification faults. 

\section{State of Work}\label{sec:consistency}
The following is a summary about the developed consistency analysis method, including the translation of TSCs into SMT problems. For reasons of space, most of the constructions can only be presented by examples. The complete constructions including correctness proofs will be given in the author's PhD thesis. 

\subsection{Defining Consistency}\label{sec:defining-consistency}
The idea behind the formal definition of a set of TSCs being consistent is not totally new. The following consistency notions adopt ideas from related work \cite{Becker2018,Becke20,FilipovikjRNS17,EllenSieverdingHungar14}. 

In general, consistency asks the question: ``Is it possible to build a system that satisfies all my requirements?'' 
Usually, this question cannot be answered unless you build the system itself. Therefore, some weaker question is used: ``Does there exist at least one trajectory (called a witness trajectory), i.e., a system observation, that does not violate one of my requirements?'' 
Obviously, if the answer to the second question is ``No'' then it is impossible to build a system that implements all the requirements together. In other words, there is some conflict in the requirements specification that makes implementation impossible and needs to be resolved. 

As a starting point for formally defining consistency of requirement TSCs we take the second question. For specifications consisting of a single TSC, we can write it as $\Sat_{\WM}(\TSC)$.  However, this can be trivially answered with $\ctrue$ for may TSCs, simply by providing a trajectory where the pre-chart is not satisfied. So, $\Sat_{\WM}(\TSC)$ alone is not an appropriate consistency criterion. 
Similar observations have been made in all of the related work \cite{Becker2018,Becke20,FilipovikjRNS17,EllenSieverdingHungar14} that considers requirements in implication form. It is solved typically by requiring that the premise of the requirement -- in TSC terms the pre-chart -- occurs at least once on the trajectory. For TSCs $\tuple{H, F, C}$, we realize this by asking for satisfiability of the basic chart
\[\HFC_\TSC = \text{\inlinetsc{
		\tscTrue\tscSeq\subchart{H}\tscSeq\tscAnd{
			\item{\subchart{F}}
			\item{\subchart{C}}
		}
}}\] with $H$, $F$, $C$ being history, future and consequence of the original TSC. Formally, it expresses the following
\begin{itemize}
	\item The pre-chart, $H$ followed by $F$, occurs at least once.
	\item The consequence $C$ is satisfied at least once in parallel to the future $F$. 
\end{itemize}
This includes only a weak approximation of the formal TSC semantics (because the consequence is checked only in parallel to \emph{one} occurrence of the future, but there may be more occurrences) but is indeed a necessary condition for what we would achieve and turns out to be sufficient to find typical specification faults\footnote{The specification fault that is observed most often is a discontinuity between consecutive spatial views in a sequence that is only satisfiable when vehicles are teleported.}. As a side effect, this removes the implicit implication between pre-chart and consequence. In related work \cite{Becke20,EllenSieverdingHungar14}, this form of consistency is called \emph{existential consistency}. 

\begin{figure*}
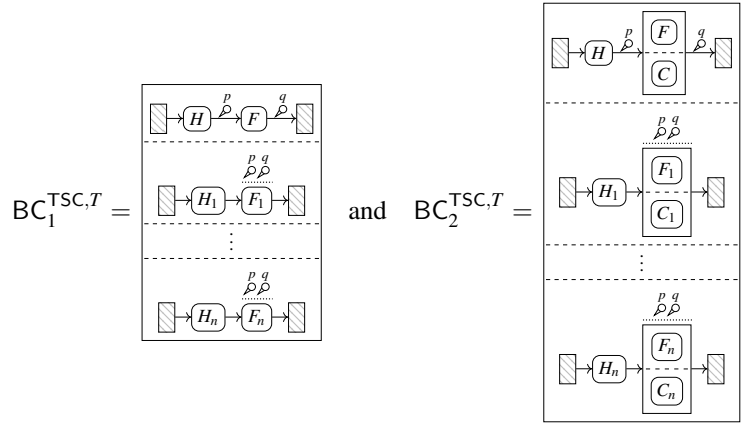

	\[
	\BC_1^{\TSC,T} = \scalebox{0.7}{\inlinetsc{
			\tscAnd{
				\item{\tscTrue\tscSeq\subchart{H}\tscXSeq{\tscPin{p}}\subchart{F}\tscXSeq{\tscPin{q}}\tscTrue}
				\item{\tscTrue\tscSeq\subchart{H_1}\tscSeq\subchart{F_1}\tscSynctxt{\tscPin{p}\tscPin{q}}\tscSeq\tscTrue}
				\item{\tscvdots}
				\item{\tscTrue\tscSeq\subchart{H_n}\tscSeq\subchart{F_n}\tscSynctxt{\tscPin{p}\tscPin{q}}\tscSeq\tscTrue}
	}}}
	\quad\text{and}\quad
	\BC_2^{\TSC,T} =\scalebox{0.7}{\inlinetsc{
			\tscAnd{
				\item{\tscTrue\tscSeq\subchart{H}\tscXSeq{\tscPin{p}}\tscAnd{\item{\subchart{F}}\item{\subchart{C}}}\tscXSeq{\tscPin{q}}\tscTrue}
				\item{\tscTrue\tscSeq\subchart{H_1}\tscSeq\tscAnd{\item{\subchart{F_1}}\item{\subchart{C_1}}}\tscSynctxt{\tscPin{p}\tscPin{q}}\tscSeq\tscTrue}
				\item{\tscvdots}
				\item{\tscTrue\tscSeq\subchart{H_n}\tscSeq\tscAnd{\item{\subchart{F_n}}\item{\subchart{C_n}}}\tscSynctxt{\tscPin{p}\tscPin{q}}\tscSeq\tscTrue}
	}}}
	\]
	\caption{Definitions of the charts $\BC_1^{\TSC,T}$ and $\BC_2^{\TSC,T}$}\label{fig:consistency-charts}
\end{figure*}

Now we lift this idea to sets (with $\text{size}>1$) of TSCs. For sets of TSCs, the witness trajectory shall show that \emph{all} TSCs are satisfied together. Recall that TSCs only constrain those time intervals on a trajectory where history and future hold. Hence, if we have witnesses for existential consistency of each TSC in a set, a witness for the whole set may be easily constructed by putting the individual witness trajectories in sequence, with some glue parts between them that are not covered by the specification. Hence, this does not give more insight on consistency of the specification than consistency of each TSC alone \cite{Becke20}. The interesting case in a set of TSCs, however, are those cases where TSCs are active in parallel. Here, consequences from different TSCs must be satisfied in parallel, which may cause actual conflicts. Of course, a specification may contain sub-sets where the pre-charts are mutual exclusive such that they, by intention cannot be active together. Therefore, the consistency notion proposed for sets of TSCs does the following.
\begin{itemize}
	\item Each sub-set of TSCs is checked separately.
	\item For each set, check if the TSCs can be active in parallel.
	\item If so, check whether also the consequences can be satisfied in parallel to the futures.
\end{itemize} 
The term \emph{active in parallel} is realized by choosing one TSC from each subset as the ``innermost'' TSC. For other TSCs, let the future start before the innermost TSC's future, and end afterwards. This way, all the TSCs are active in parallel. For some innermost TSC $\TSC=\tuple{H, F, C}$ and a context $T = \{\tuple{H_1, F_1, C_1}, \dots, \tuple{H_n, F_n, C_n}\}$ this is encoded in the basic charts $\BC_1^{\TSC,T}$ and $\BC_2^{\TSC,T}$ shown in Figure~\ref{fig:consistency-charts},
where $\BC_1^{\TSC,T}$ does include the consequences and $\BC_2^{\TSC,T}$ does not. 
Because timing constraints may restrict the duration of the future, it is for some TSCs not possible to be the innermost TSC. Therefore, every TSC is tried as the innermost one. 

As already mentioned before, we don't try to solve the satisfiability problem exactly. Instead, we approach it from both directions building two semi-decision procedures $\CheckSatN_{\WM}$ and $\CheckSatS_{\WM}$. They check a \underline{n}eccessary respectively \underline{s}ufficient condition for satisfiability of a basic chart in context of our world model $\WM$. 
Taking some basic chart $\BC$ as input, the two functions each do the following. 
\begin{enumerate}
	\item Derive a BMC problem from $\BC$ and $\WM$, and an unrolling depth $n$.
	\item Unroll the BMC problem for $n$ steps yielding a constraint $\Phi$
	\item Send $\Phi$ to an SMT solver and return the result. 
\end{enumerate}
The functions differ in the constructed BMC problem and chosen unrolling depth and are explained in more detail in the following sections. $\CheckSatN_\WM(\BC)$ returns \texttt{unsat} only if $\BC$ is \emph{unsatisfiable} and $\CheckSatS_\WM(\BC)$ returns \texttt{sat} only if $\BC$ is \emph{satisfiable} within $\WM$. 

Algorithm~\ref{alg:partial-consistency} lists the procedure for finding inconsistencies in a set $\TSC$ using both approximations. Recall that a single TSC is existentially inconsistent if $\CheckSatN_{\WM}\left(\HFC_\TSC\right)=\texttt{unsat}$. In the case $T=\emptyset$ is $\BC_2^{\TSC,T}$ equivalent to $\HFC_\TSC$ above. Hence, the algorithm reports also existetial inconsistencies of single TSCs in the set. Note also that $\CheckSatN$ and $\CheckSatS$ are chained in a way that an inconsistency is only reported if an exact decision procedure would yield the same result. Provided that the implementation is sound, the approximations cannot result in spuriously reported inconsistencies.  

\begin{algorithm}[t]
	\ForEach{$\TSC\in\TSCs$}{
		\ForEach{$T \subseteq \TSCs\setminus \{\TSC\}$}{
			\If{$T=\emptyset$ \Or $\CheckSatS_{\WM}\left(\BC_1^{\TSC,T}\right)=\texttt{sat}$}{
				\If{ $\CheckSatN_{\WM}\left(\BC_2^{\TSC,T}\right)=\texttt{unsat}$}{
					Report inconsistency of $\{\TSC\}\cup T$\;
				}
			}
		}
	}
	\caption{Finding inconsistencies in TSCs}\label{alg:partial-consistency}
\end{algorithm}

The algorithm has much room for optimization. 
\begin{itemize}
	\item Usually, the user is interested in the \emph{minimum inconsistent subsets}, i.e., inconsistent sets that become consistent when removing one TSC. So we can skip any subset where we know that it contains an inconsistent subset. 
	\item For sets $S \supseteq T$, unsatisfiability of $\BC_1^{\TSC,T}$ implies unsatisfiablility of $\BC_1^{\TSC,S}$ (the same holds for the corresponding BMC problems). So we can skip any superset of $T$ if $\CheckSatS_{\WM}\left(\BC_1^{\TSC,T}\right)=\texttt{unsat}$ in line 3 of the algorithm. 
	\item The BMC problem generated by $\CheckSatS$ is usually harder to solve than the one from $\CheckSatN$. so we check it only if $\CheckSatN_{\WM}\left(\BC_2^{\TSC,T}\right)=\texttt{unsat}$. 
\end{itemize}
The prototype used for evaluation implements these optimizations, and experiments (see Section~\ref{sec:evaluation}) show that most of the cases can be skipped. 

\subsection{Semi-deciding Satisfiability of Basic Charts}\label{sec:decision-procedure}
Both $\CheckSatS$ and $\CheckSatN$ work by encoding a basic chart as a BMC problem, unrolling it for $n$ steps, and running an SMT solver on it. As a necessary condition for satisfiability, we use SMT solving to find a satisfying witness trajectory. The same technique has been used in related work  \cite{Becke20} for simulation of TSCs. The same idea is also followed for the necessary condition. The differences are described later on in Section~\ref{sec:CheckSatN}. In both cases, the BMC problem $(I, T, F)$ is of the form
\begin{align*} 
I & \equiv I_{\text{chart}} \\
T & \equiv T_{\text{chart}} \wedge T_{\WM} \wedge \bigwedge_{sv\in SV} (b_{sv} \leftrightarrow T_{sv}) \\
F &\equiv F_{\text{chart}}  
\end{align*}
The triple $(I_{\text{chart}}, T_{\text{chart}}, F_{\text{chart}})$ encodes the chart structure, where satisfaction of a spatial view $sv$ in the current step is substituted by some variable $b_{sv}$. The actual formula for $sv$ (ranging over all spatial views in the chart) is encoded in the constraint $T_{sv}$. Finally, $T_\WM$ encodes world model constraints, i.e. it restricts solutions to trajectories from the world model. 

\subsubsection{Encoding the Chart Structure}
Encoding of the chart structure is the same for both sufficient and necessary conditions, except that $\CheckSatS$ uses a fixed time step\footnote{Because of efficiency and decidability, we restrict ourselves to linear inequalities only. With variable time steps, the relation between velocity and position becomes non-linear.} and $\CheckSatN$ variable step size. Start and end of sub-charts (with index $i$) is encoded with Boolean variables $\started_i$ and $\complete_i$. 

\begin{example}
	Encoding the chart structure of
	\inlinetsc{\tscTrue\tscSeq\tscInvariant{$A$}\tscSeq\tscTrue} 
	leads to the constraints shown in Table~\ref{tab:example-chart-constraints}. The invariant nodes are numbered from left to right. Note that no variable $\started_1$ is needed for the first sub-chart, because it always starts with the trajectory. The variable $\ok_2$ keeps track of the second invariant after start of the sub-chart (for empty spatial views this is not needed because they cannot be violated).   
\end{example}
Another example can be found in the related work \cite{Becke20}. 

\begin{table*}[tb]
\caption[]{The chart structure of \inlinetsc{\tscTrue\tscSeq\tscInvariant{$A$}\tscSeq\tscTrue} encoded in a BMC problem.}\label{tab:example-chart-constraints}
\[
\begin{array}{l@{\qquad}l@{\qquad}l}
	\hline
	\text{Initial} & \text{Transition} & \text{Final} \\
	\hline\hline
	\neg\complete_1 & \started'_2 \rightarrow (\complete_1 \vee \started_2)  & \complete_3\\
	\neg \complete_2 & \complete'_2 \leftrightarrow (\started_2 \wedge \ok'_2) & \\
	\ok_2 & \ok'_2 \leftrightarrow (\started_2 \rightarrow \ok_2 \wedge b_A) &  \\
	\neg \complete_3 & \started'_3 \rightarrow (\complete'_2 \vee \started_3) &  \\
	& \complete'_3 \leftrightarrow \started_3 & \\
	\hline
\end{array}
\]
\end{table*}

\subsubsection{Sufficient Conditions for Invariants}\label{sec:CheckSatS}
Now, we have a closer look to the spatial view constraints $\Phi_{sv}$ and world model constraints $\Phi_{\WM}$. Attribute values of non-dynamic objects (such as lanes) are represented by free variables in the BMC problem. So, for some lane $\mathtt{l}$ we'd introduce four variables $\mathtt{y_l}$, $\mathtt{start_l}$, and $\mathtt{end_l}$ -- derived attributes can be inlined and don't need variables. The modeling of cars is more complicated and differs for $\CheckSatS$ and $\CheckSatN$. In the following, $\CheckSatS$ is described.   

The trajectories of cars are described as quadratic Bézier splines.  Each segment of the spline is described by control points 
\[\mathbf{p}_0 = (x_0, y_0),\qquad\mathbf{p}_1 = (x_1, y_1),\qquad\mathbf{p}_2=\mathbf{p}_0'=(x_0', y_0')\enspace.\] 
Start and end point of consecutive segments are shared. Each segment describes the movement within one unrolling step of the BMC problem. The position at time $t$ after start of the current step (with step size $\Delta$) is given by the function
\[\mathbf{p}(t)=\mathbf{p}_0\Big(1-\frac{t}{\Delta}\Big)^2 + 2\mathbf{p}_1\Big(1-\frac{t}{\Delta}\Big)\frac{t}{\Delta} + \mathbf{p}_2\frac{t^2}{\Delta^2}\] and the velocity vector is determined by its derivative. Other attributes, such as the bounding box extends, are encoded by two variables standing for safe upper and lower bounds in the current step. The special properties of Bézier splines (see \cite{Prautzsch2013} for an overview) allow a quite convenient encoding of spatial views. For example, distances between Bézier splines can be expressed as pairwise distances between their control points. Furthermore, every Bézier spline segment lies within the convex polygon spanned by its control points. These properties make it possible to conservatively approximate spatial views by linear constraints over the control points.  
\begin{example}
In $\CheckSatS$, the spatial view in Figure~\ref{fig:sv-distance} is encoded as 
\begin{align*}
	\bigwedge_{i \in \{0,1\}} & \left(\begin{array}{l}
		x_{\carJ,i} + \mathtt{bbxmin}_{\carJ}^l \\ \qquad - x_{\carI,i} - \mathtt{bbxmax}_{\carI}^u > 5m
	\end{array}\right)\\
	\wedge~ & \left(\begin{array}{l}
		x_{\carJ,0}' + \mathtt{bbxmin}_{\carJ}^l \\\qquad - x_{\carI,0}' - \mathtt{bbxmax}_{\carI}^u \geq 5m
	\end{array}\right)
\end{align*}
\end{example}
If this formula holds, then the spatial view is satisfied on the whole spline segment. 

The world model constraints $\Phi_\WM$ ensure that $\mathtt{bbxmin}^l$ and $\mathtt{bbxmax}^u$ are safe upper and lower bounds. Because they depend on the heading angle $\theta$, we choose a piece-wise approximation where different heading angles are approximated by a finite set $\mathcal{I}$ of intervals $I \in \mathcal{I}$. E.g., $\mathtt{bbxmax}$ of $\carI$ is characterized by a constraint
\begin{multline*}\label{eq:bbox}
	\bigvee_{I \in \mathcal{I}} \phi_{\carI.\theta \in I} \wedge \Big(\mathtt{bbxmax}^l \leq \inf_{\theta \in I} \{\mathtt{bbxmax}(\theta)\} \\
	\wedge \mathtt{bbxmax}^u \geq \sup_{\theta \in I} \{\mathtt{bbxmax}(\theta)\}\Big)
\end{multline*}
where $\phi_{\carI.\theta \in I}$ is a linear constraint describing that the heading is within $I$. The infimum and supremum of the heading-dependent bounding box extend $\mathtt{bbxmax}(\theta)$ can be calculated numerically when generating the BMC problem. Furthermore, constrains are added that ensure that the trajectory is continuously differentiable with curvature $|\kappa| \leq \frac{\tan\delta_{\max}}{L}$ and lateral acceleration $|a_{lat}| = |\mathtt{v}^2\kappa| = |(\dot{\mathtt{x}}^2 + \dot{\mathtt{y}}^2)\kappa| \leq 0.4g$. These ensure that generated trajectories are solutions of the single track model specified in Section~\ref{sec:worldmodel}.  

\subsubsection{Necessary Conditions for Invariants}\label{sec:CheckSatN}
For $\CheckSatN$, the chart structure is encoded the same way in the BMC problem, but the constraints $T_{\WM}$ and $T_{sv}$ are relaxed. Following a result by Fränzle and Hansen about positive Duration Calculus \cite{FraenzleH07}, we know that the chart structure BMC problem is either unsatisfiable or has a solution after $m+1$ unrolling steps, with $m$ being the number of sequence operation in the TSC. However, in a solution with minimum unrolling depth, a single BMC step may cover an arbitrarily long time interval. For example, during one step, a whole overtaking maneuver may take place. Therefore, instead of describing and restricting the exact vehicle movement, invariants are  checked at discrete time points only. 

\begin{example}
	In $\CheckSatN$, the spatial view in Figure~\ref{fig:sv-distance} is encoded as
	\begin{align*}
	& x_{\carJ} + \mathtt{bbxmin}_{\carJ} - x_{\carI} - \mathtt{bbxmax}_{\carI} > 5m \\
	{}\wedge{} & x_{\carJ}' + \mathtt{bbxmin}_{\carJ}' - x_{\carI}' - \mathtt{bbxmax}_{\carI}' \geq 5m
	\end{align*}
	where variables stand for object attribute values in the current, respective next, step. 
\end{example}
By checking spatial views (as in the above formula) both at beginning and end of a step, we catch both contradictions between parallel as well as consecutive invariant nodes. Note that, because the end time $e$ is excluded in Definition~\ref{def:chart-semantics} when evaluating invariant nodes, we must only enforce non-strict spatial relations for the end of a step. 
Because we don't need to characterize the evaluation of the bounding box for the whole unrolling step anymore, the upper and lower bounds can be replaced by single variables. 

	\subsection{Evaluation}\label{sec:evaluation}

To avoid that the results are biased by the author's intention when designing the use case, the evaluation is based on an earlier case study \cite{Esterle2021}. In this work, common highway traffic rules have been formalized in LTL. In the original work \cite{Esterle}, they have been used to examine databases of recorded traffic data for rule violations. In order to translate them to TSCs, first spatial views have been designed that match the atomic propositions in the LTL rules. An example is the proposition $\textit{\textsf{succ}}^{i\rightarrowtail j}$ saying $\carI$ is the successor of $\carJ$ shown in Figure~\ref{fig:sv-nowhere}. Then, the LTL expressions have been translated to TSCs using a set of pre-defined patterns. The result is a set of nine TSCs -- rules that do not apply to two-lane highways have been omitted.  

\begin{figure*} 
	\centering
	\subcaptionbox{keep in right-most lane}{\includegraphics[scale=.8]{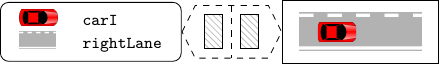}} \\[1em]
	\subcaptionbox{no passing on the right side}{\includegraphics[width=\textwidth]{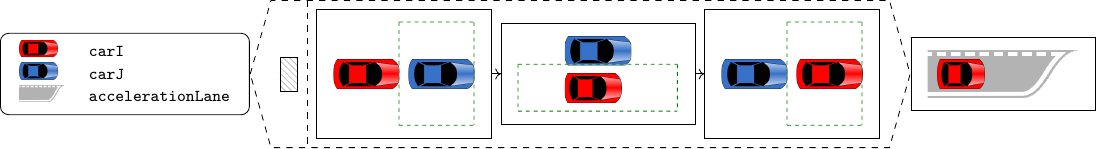}} \\[1em]
	\subcaptionbox{safe lane change}{\includegraphics[scale=.8]{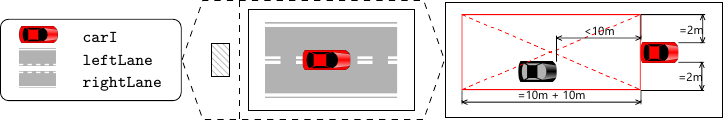}} 
	\caption{Conflicting traffic rules found in the evaluation}\label{fig:ltl-tscs}
\end{figure*}

The analysis finds that the TSCs shown in Figure~\ref{fig:ltl-tscs} are pairwise inconsistent. Taking a closer look at the TSCs shows that TSC (a) forbids to use any other lane than the right one, making lane-change rules (TSCs (b) and (c)) inapplicable. The conflict between (b) and (c) comes from the chosen formalization of (b) as a TSC: the forbidden behavior (right overtaking) is in the pre-chart, and the exception (being on an acceleration lane) to this rule forms the consequence. From the TSC semantics point of view, this is correct, but breaks the intuition behind premise-consequence charts. 

The BMC problems for $\CheckSatS$ have been configured with a step size of \unit{3}{\second} and unrolled for 10 steps\footnote{The unrolling depth for $\CheckSatN$ is calculated automatically}. From the 2304 cases that need to be checked in the worst case when analyzing consistency of nine TSCs, only 467 (\unit{21}{\%}) must be handed over to the SMT solver; the rest could be inferred from existing results using the optimizations mentioned at the end of Section~\ref{sec:defining-consistency}. In total, the analysis takes about \unit{1}{\minute} and \unit{20}{\second} on a Windows 10 notebook with an Intel Core i7-4700MQ CPU @ \unit{2.39}{\giga\hertz} and \unit{8}{\giga\byte} RAM, using version 4.6.0 of the Z3 solver \cite{MouraB08} with default settings. The results show that in practice only small subsets (up to size 3 in the experiment) of TSCs need to be checked, so there is only a polynomial growth in the number of cases, and the individual SMT problems remain manageably small. In \cite{Becker2022}, it has been shown that also more complex TSCs can be checked for satisfiability in reasonable time. 

A similar experiment has been carried out on a specification done by a non-expert in formal specification\footnote{A student employee at the author's institute who was not under the author's supervision. }. This second experiment shows that the consistency analysis is especially strong in finding formalization errors such as unsatisfiable spatial constraints or impossible sequences. The scalability of the TSC encoding has been evaluated in the context of already published work \cite{Becker2022}. 

\subsection{Practical Application}

\begin{figure}
	\includegraphics[width=\columnwidth]{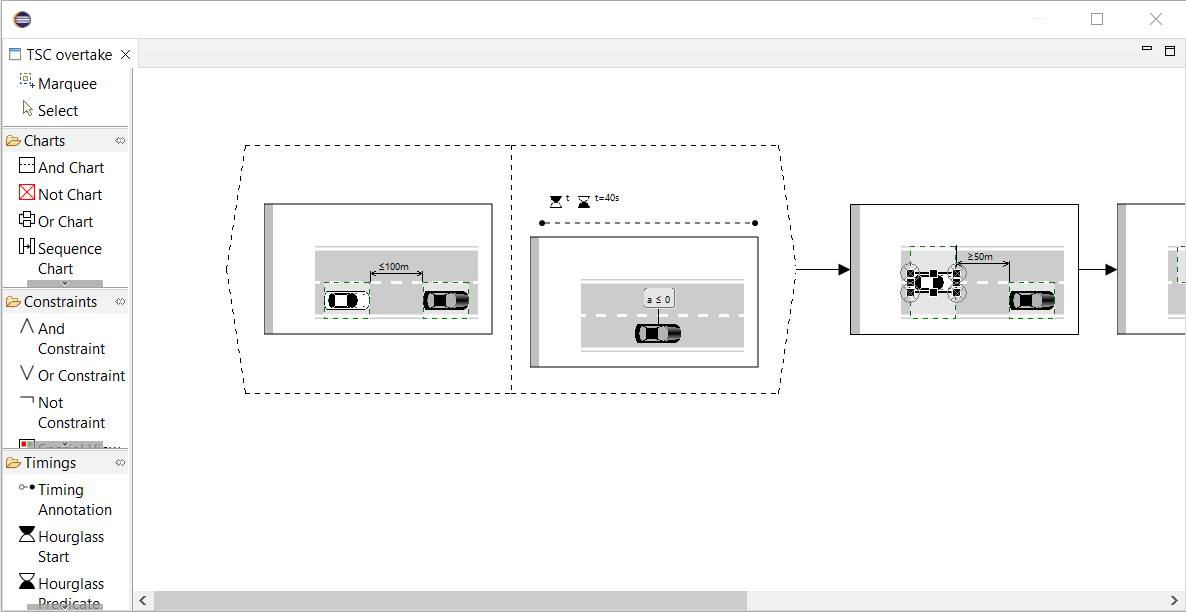}
	\caption{Screenshot of the graphical specification tool for TSCs}\label{fig:tsc-editor}
\end{figure}

The consistency analysis prototype has been integrated into a graphical specification tool for TSCs (Figure~\ref{fig:tsc-editor}). It is part of a tool chain for scenario-based development that is under ongoing development at the DLR SE institute. Other parts of the tool chain include generation of OpenSCENARIO files \cite{Becker2022} and scenario monitors \cite{Grundt2022}. The tooling is actively used both internally and in cooperation with industry partners. Furthermore, tooling and methods are extended to the railway and maritime transport domains as well. The future use of the specification tooling will yield more insight about practical suitability of the consistency analysis. 

	\section{Conclusion and Outlook}\label{sec:conclusion}

This work proposes a consistency analysis technique for TSCs. Although being grounded on the formal semantics of TSCs, the goal is not to produce a certificate of formal requirements consistency, but to find real conflicts in the specification on a well-defined basis. Consistency is reduced to satisfiability of existential TSCs, which in turn is encoded as a BMC problem. For reason of space, the latter has only been sketched here with some examples. The full construction, including correctness proofs, will be given in the author's PhD thesis. It shall be highlighted that the implementation (that has been used to carry out the presented evaluation) assumes the shown parameterized vehicle model, but otherwise is not bound to a specific domain ontology. 

The proposed consistency analysis re-uses basic ideas from earlier work \cite{Becke20}. The consistency notion for single TSCs remains the same. The idea for consistency of non-singleton sets is also similar. Both approaches split analysis cases each into a premise (can the TSCs be active in parallel?) and a consequence (is there a trajectory where TSCs are active in parallel?). However, the technical realization of the term ``active in parallel'' has been reworked. The earlier work synchronizes invariant nodes from pairs of TSCs. Although it is possible to extend this to larger subsets, these number of analysis cases per subset explodes. In contrast, the current approach work only requires $n$ cases per subset of size $n$, and it is possible to re-use information and skip some of the cases. Furthermore, the earlier work uses a necessary condition for both the premise and the consequence. So, there is no guarantee that false inconsistencies are ruled out. In the present work, this is solved by generating a satisfying trajectory \cite{Becker2022} instead of an approximation thereof. The analysis tool then creates a trajectory for every (minimal) inconsistent subset that shows how a conflicting situation arises. By the use of a Bézier spline encoding technique with additional constraints (see Section~\ref{sec:CheckSatS}), the generated trajectories are solutions of a single-track vehicle dynamics model. Simpler (e.g., piece-wise linear) trajectories would not have this property, while alternative trajectory encodings (such as arc segments or clothoids) can be hardly expressed by linear constraint systems. Related work \cite{Becke20} shows that the generated trajectories can be simulated. 

The experimental results demonstrate practical applicability of the approach. Furthermore, it turns out that it scales in practice because constraint solving is only needed for small subsets. However, it would be worthwhile to support the experimental results with further case studies. So far, the present case study shows that the consistency analysis catches both contradicting as well as ill-structured (violating specification patterns) requirements. Therefore, it is future work to develop explicit specification guidelines for the work with TSCs, and to extend the consistency analysis to handle a broader set of specification patterns. 

Besides the application in the presented consistency analysis method, parts of the used techniques (and their implementation) are planed to be re-used in other applications of TSCs. For example, trajectory generation can be used for simulation in the context of criticality analysis, as well as test case generation. Translation of spatial invariants to constraint systems is also required for monitoring.  

	\bibliographystyle{apalike}
	{\small
	\bibliography{bibliography}}

\begin{thebibliography}{}

\bibitem[Aichernig et~al., 2015]{AichernigHLNT15}
Aichernig, B.~K., H{\"{o}}rmaier, K., Lorber, F., Ni{\v{c}}kovi{\'{c}}, D., and
  Tiran, S. (2015).
\newblock Require, test and trace {IT}.
\newblock In {\em Formal Methods for Industrial Critical Systems - 20th
  International Workshop, {FMICS} 2015, Proceedings}, volume 9128 of {\em
  LNCS}, pages 113--127. Springer.

\bibitem[Armando et~al., 2009]{Armando2009}
Armando, A., Mantovani, J., and Platania, L. (2009).
\newblock Bounded model checking of software using smt solvers instead of sat
  solvers.
\newblock {\em International Journal on Software Tools for Technology
  Transfer}, 11:69--83.

\bibitem[Baader et~al., 2007]{BaaderCMNP07}
Baader, F., Calvanese, D., Mcguinness, D., Nardi, D., and Patel-Schneider, P.
  (2007).
\newblock {\em The Description Logic Handbook: Theory, Implementation, and
  Applications}.

\bibitem[Barrett et~al., 2009]{BarrettSST09}
Barrett, C.~W., Sebastiani, R., Seshia, S.~A., and Tinelli, C. (2009).
\newblock Satisfiability modulo theories.
\newblock {\em Handbook of satisfiability}, 185:825--885.

\bibitem[Becker et~al., 2022]{Becker2022}
Becker, J., Koopmann, T., Neurohr, B., Neurohr, C., Westhofen, L., Wirtz, B.,
  B{\"o}de, E., and Damm, W. (2022).
\newblock Simulation of {{Abstract Scenarios}}: {{Towards Automated Tooling}}
  in {{Criticality Analysis}}.
\newblock pages 42--51.

\bibitem[Becker, 2018]{Becker2018}
Becker, J.~S. (2018).
\newblock Analyzing consistency of formal requirements.
\newblock In {\em 18th International Workshop on Automated Verification of
  Critical Systems}.

\bibitem[Becker, 2020]{Becke20}
Becker, J.~S. (2020).
\newblock Partial consistency for requirement engineering with traffic sequence
  charts.
\newblock In {\em Automotive Software Engineering (ASE2020)}.

\bibitem[Becker, 2024]{Becker2024}
Becker, J.~S. (2024).
\newblock Safe linear encoding of vehicle dynamics for the instantiation of
  abstract scenarios.
\newblock In {\em Formal Methods for Industrial Critical Systems}, volume 14952
  of {\em LNCS}. Springer.

\bibitem[Bender et~al., 2014]{bender2014lanelets}
Bender, P., Ziegler, J., and Stiller, C. (2014).
\newblock Lanelets: Efficient map representation for autonomous driving.
\newblock In {\em 2014 IEEE Intelligent Vehicles Symposium Proceedings}, pages
  420--425. IEEE.

\bibitem[Booch, 2005]{Booch2005}
Booch, G. (2005).
\newblock {\em The unified modeling language user guide}.
\newblock Pearson Education India.

\bibitem[Bouajjani et~al., 1995]{Bouajjani1995}
Bouajjani, A., Lakhnech, Y., and Robbana, R. (1995).
\newblock From duration calculus to linear hybrid automata.
\newblock In {\em {LNCS}}, volume 939, pages 196--210.

\bibitem[Chaochen et~al., 1993]{ChaochenHS93}
Chaochen, Z., Hansen, M.~R., and Sestoft, P. (1993).
\newblock Decidability and undecidability results for duration calculus.
\newblock In {\em {STACS} 93, 10th Annual Symposium on Theoretical Aspects of
  Computer Science, W{\"{u}}rzburg, Germany, February 25-27, 1993,
  Proceedings}, pages 58--68.

\bibitem[Chen and Sankaranarayanan, 2016]{chen2016}
Chen, X. and Sankaranarayanan, S. (2016).
\newblock Decomposed reachability analysis for nonlinear systems.
\newblock In {\em Real-Time Systems Symposium (RTSS), 2016 IEEE}, pages 13--24.
  IEEE.

\bibitem[Damm et~al., 2017]{DammKMPR17}
Damm, W., Kemper, S., M{\"o}hlmann, E., Peikenkamp, T., and Rakow, A. (2017).
\newblock Traffic sequence charts - from visualization to semantics.
\newblock Reports of SFB/TR 14 AVACS 117, SFB/TR 14 AVACS.

\bibitem[Damm et~al., 2018a]{Damm2018}
Damm, W., Kemper, S., M{\"o}hlmann, E., Peikenkamp, T., and Rakow, A. (2018a).
\newblock {Using Traffic Sequence Charts for the Development of HAVs}.
\newblock In {\em {ERTS 2018}}, 9th European Congress on Embedded Real Time
  Software and Systems (ERTS 2018), Toulouse, France.

\bibitem[Damm et~al., 2018b]{DammMPR18}
Damm, W., M{\"o}hlmann, E., Peikenkamp, T., and Rakow, A. (2018b).
\newblock {\em A Formal Semantics for Traffic Sequence Charts}, pages 182--205.
\newblock Springer International Publishing, Cham.

\bibitem[Damm et~al., 2019]{DammMR19a}
Damm, W., M{\"o}hlmann, E., and Rakow, A. (2019).
\newblock A scenario discovery process based on traffic sequence charts.
\newblock In {\em Validation {\&} Verification of Automated Systems -- Results
  of the {ENABLE-S3} Project}.

\bibitem[de~Moura and Bj{\o}rner, 2008]{MouraB08}
de~Moura, L. and Bj{\o}rner, N. (2008).
\newblock {Z3}: An efficient {SMT} solver.
\newblock In {\em Tools and Algorithms for the Construction and Analysis of
  Systems: 14th International Conference, TACAS 2008. Proceedings}, pages
  337--340. Springer.

\bibitem[Dwyer et~al., 1999]{dwyer1999patterns}
Dwyer, M.~B., Avrunin, G.~S., and Corbett, J.~C. (1999).
\newblock Patterns in property specifications for finite-state verification.
\newblock In {\em Proceedings of the 21st international conference on Software
  engineering}, pages 411--420.

\bibitem[Eggers et~al., 2011]{eggers2011}
Eggers, A., Ramdani, N., Nedialkov, N., and Fr{\"a}nzle, M. (2011).
\newblock Improving sat modulo ode for hybrid systems analysis by combining
  different enclosure methods.
\newblock In {\em International Conference on Software Engineering and Formal
  Methods}, pages 172--187. Springer.

\bibitem[Eggers et~al., 2018]{EggerSTBB18}
Eggers, A., Stasch, M., Teige, T., Bienmüller, T., and Brockmeyer, U. (2018).
\newblock Constraint systems from traffic scenarios for the validation of
  autonomous driving.
\newblock In {\em Third International Workshop on Satisfiability Checking and
  Symbolic Computation, Part of FLOC 2018}.

\bibitem[Ellen et~al., 2014]{EllenSieverdingHungar14}
Ellen, C., Sieverding, S., and Hungar, H. (2014).
\newblock Detecting consistencies and inconsistencies of pattern-based
  functional requirements.
\newblock In Lang, F. and Flammini, F., editors, {\em Formal Methods for
  Industrial Critical Systems - 19th International Conference, {FMICS} 2014,
  Florence, Italy, September 11-12, 2014. Proceedings}, volume 8718 of {\em
  Lecture Notes in Computer Science}, pages 155--169. Springer.

\bibitem[Esterle, 2021]{Esterle2021}
Esterle, K. (2021).
\newblock {\em Formalizing and Modeling Traffic Rules Within Interactive
  Behavior Planning}.
\newblock PhD thesis, Universit{\"a}t M{\"u}nchen.

\bibitem[Esterle et~al., 2020]{Esterle}
Esterle, K., Gressenbuch, L., and Knoll, A. (2020).
\newblock Formalizing traffic rules for machine interpretability.
\newblock In {\em 2020 IEEE 3rd Connected and Automated Vehicles Symposium
  (CAVS)}, pages 1--7. IEEE.

\bibitem[Feiler et~al., 2010]{Feiler2010}
Feiler, P., Wrage, L., and Hansson, J. (2010).
\newblock System architecture virtual integration: A case study.
\newblock In {\em Embedded Real-time Software and Systems Conference}.

\bibitem[Filipovikj et~al., 2017]{FilipovikjRNS17}
Filipovikj, P., Rodriguez-Navas, G., Nyberg, M., and Seceleanu, C. (2017).
\newblock Smt-based consistency analysis of industrial systems requirements.
\newblock In {\em Proceedings of the Symposium on Applied Computing}, pages
  1272--1279. ACM.

\bibitem[Fr{\"a}nzle and Hansen, 2007]{FraenzleH07}
Fr{\"a}nzle, M. and Hansen, M.~R. (2007).
\newblock Deciding an interval logic with accumulated durations.
\newblock In {\em International Conference on Tools and Algorithms for the
  Construction and Analysis of Systems}, volume 4424 of {\em LNCS}, pages
  201--215. Springer.

\bibitem[Frehse, 2005]{frehse2005phaver}
Frehse, G. (2005).
\newblock Phaver: Algorithmic verification of hybrid systems past hytech.
\newblock In {\em International workshop on hybrid systems: computation and
  control}, pages 258--273. Springer.

\bibitem[Frehse et~al., 2011]{frehse2011spaceex}
Frehse, G., Le~Guernic, C., Donz{\'e}, A., Cotton, S., Ray, R., Lebeltel, O.,
  Ripado, R., Girard, A., Dang, T., and Maler, O. (2011).
\newblock Spaceex: Scalable verification of hybrid systems.
\newblock In {\em International Conference on Computer Aided Verification},
  pages 379--395. Springer.

\bibitem[Grundt et~al., 2022]{Grundt2022}
Grundt, D., K{\"o}hne, A., Saxena, I., Stemmer, R., Westphal, B., and
  M{\"o}hlmann, E. (2022).
\newblock Towards runtime monitoring of complex system requirements for
  autonomous driving functions.

\bibitem[Heitmeyer et~al., 1996]{heitmeyer1996automated}
Heitmeyer, C.~L., Jeffords, R.~D., and Labaw, B.~G. (1996).
\newblock Automated consistency checking of requirements specifications.
\newblock {\em ACM Transactions on Software Engineering and Methodology
  (TOSEM)}, 5(3):231--261.

\bibitem[Henzinger et~al., 1997]{henzinger1997hytech}
Henzinger, T.~A., Ho, P.-H., and Wong-Toi, H. (1997).
\newblock Hytech: A model checker for hybrid systems.
\newblock {\em International Journal on Software Tools for Technology
  Transfer}, 1(1-2):110--122.

\bibitem[{ISO~26262}, 2018]{ISO26262}
{ISO~26262} (2018).
\newblock Road vhicles---functional safety.
\newblock Technical report, International Organisation for Standardization.

\bibitem[Jaffe et~al., 1991]{jaffe1991}
Jaffe, M.~S., Leveson, N.~G., Heimdahl, M. P.~E., and Melhart, B.~E. (1991).
\newblock Software requirements analysis for real-time process-control systems.
\newblock {\em IEEE transactions on software engineering}, 17(3):241--258.

\bibitem[Kalisvaart et~al., 2020]{Kalisvaart2020}
Kalisvaart, S., Slavik, Z., and {Op den Camp}, O. (2020).
\newblock Using scenarios in safety validation of automated systems.
\newblock In Leitner, A., Watzenig, D., and Ibanez-Guzman, J., editors, {\em
  Validation and Verification of Automated Systems}, pages 27--44. {Springer
  International Publishing}, Cham.

\bibitem[Kamalrudin and Sidek, 2015]{kamalrudin2015review}
Kamalrudin, M. and Sidek, S. (2015).
\newblock A review on software requirements validation and consistency
  management.
\newblock {\em International journal of software engineering and its
  applications}, 9(10):39--58.

\bibitem[Konrad and Cheng, 2005]{konrad2005real}
Konrad, S. and Cheng, B.~H. (2005).
\newblock Real-time specification patterns.
\newblock In {\em Proceedings of the 27th international conference on Software
  engineering}, pages 372--381.

\bibitem[Koopman and Fratrik, 2019]{koopman2019many}
Koopman, P. and Fratrik, F. (2019).
\newblock How many operational design domains, objects, and events?
\newblock In {\em Safeai@ aaai}.

\bibitem[Kramer et~al., 2020]{Kramer2020}
Kramer, B., Neurohr, C., B{\"u}ker, M., B{\"o}de, E., Fr{\"a}nzle, M., and
  Damm, W. (2020).
\newblock Identification and quantification of hazardous scenarios for
  automated driving.
\newblock In Zeller, M. and H{\"o}fig, K., editors, {\em Model-Based Safety and
  Assessment}, pages 163--178, Cham. Springer International Publishing.

\bibitem[Leitner et~al., 2019]{ENABLES3SummaryResults2019}
Leitner, A. et~al. (2019).
\newblock {{ENABLE-S3 Summary}} of {{Results}}.
\newblock Technical report.

\bibitem[Leveson, 2000]{leveson2000completeness}
Leveson, N. (2000).
\newblock Completeness in formal specification language design for
  process-control systems.
\newblock In {\em Proceedings of the third workshop on Formal methods in
  software practice}, pages 75--87.

\bibitem[Menzel et~al., 2018]{Menzel2018}
Menzel, T., Bagschik, G., and Maurer, M. (2018).
\newblock Scenarios for development, test and validation of automated vehicles.
\newblock In {\em 2018 IEEE Intelligent Vehicles Symposium (IV)}, pages
  1821--1827. IEEE.

\bibitem[Neurohr et~al., 2021]{Neurohr2021}
Neurohr, C., Westhofen, L., Butz, M., Bollmann, M.~H., Eberle, U., and Galbas,
  R. (2021).
\newblock Criticality analysis for the verification and validation of automated
  vehicles.
\newblock {\em IEEE Access}, 9:18016--18041.

\bibitem[Plaku et~al., 2007]{plaku2007hybrid}
Plaku, E., Kavraki, L.~E., and Vardi, M.~Y. (2007).
\newblock Hybrid systems: From verification to falsification.
\newblock In {\em International Conference on Computer Aided Verification},
  pages 463--476. Springer.

\bibitem[Plaku et~al., 2013]{plaku2013falsification}
Plaku, E., Kavraki, L.~E., and Vardi, M.~Y. (2013).
\newblock Falsification of ltl safety properties in hybrid systems.
\newblock {\em International Journal on Software Tools for Technology
  Transfer}, 15(4):305--320.

\bibitem[Post et~al., 2011]{PostHP11}
Post, A., Hoenicke, J., and Podelski, A. (2011).
\newblock rt-inconsistency: {A} new property for real-time requirements.
\newblock In {\em Fundamental Approaches to Software Engineering - 14th
  International Conference, {FASE} 2011. Proceedings}, pages 34--49.

\bibitem[Prautzsch et~al., 2013]{Prautzsch2013}
Prautzsch, H., Boehm, W., and Paluszny, M. (2013).
\newblock {\em B{\'e}zier and B-spline techniques}.
\newblock Springer Science \& Business Media.

\bibitem[Riedmaier et~al., 2020]{Riedmaier2020}
Riedmaier, S., Ponn, T., Ludwig, D., Schick, B., and Diermeyer, F. (2020).
\newblock Survey on scenario-based safety assessment of automated vehicles.
\newblock {\em IEEE Access}, 8:87456--87477.

\bibitem[{SAE~J3016\_202104}, 2021]{SAE3016}
{SAE~J3016\_202104} (2021).
\newblock Taxonomy and definitions for terms related to driving automation
  systems for on-road motor vehicles.
\newblock Technical report, International Organisation for Standardization.

\bibitem[Schramm et~al., 2014]{Schramm2014}
Schramm, D., Hiller, M., and Bardini, R. (2014).
\newblock Single track models.
\newblock In {\em Vehicle Dynamics}, pages 223--253. Springer.

\bibitem[Zowghi and Gervasi, 2003]{ZowghiGervasi03}
Zowghi, D. and Gervasi, V. (2003).
\newblock On the interplay between consistency, completeness, and correctness
  in requirements evolution.
\newblock {\em Information {\&} Software Technology}, 45(14):993--1009.

\end{thebibliography}
\end{document}